\documentclass{jps-cp}
\usepackage{txfonts} 
\usepackage{url}
\usepackage{graphicx}

\title{Impact of Uncertainties in Nuclear Reaction Cross Sections on $p$ Nucleosynthesis in Thermonuclear Supernovae}

\author{T. \textsc{Rauscher}$^{1,2,3}$, N. \textsc{Nishimura}$^{4}$, G. \textsc{Cescutti}$^{5}$, R. \textsc{Hirschi}$^{3,6,7}$,
A. St.J. \textsc{Murphy}$^{3,8}$,\\
and C. \textsc{Travaglio}$^{9}$}

\inst{$^{1}$Department of Physics, University of Basel, 4056 Basel, Switzerland\\
$^{2}$Centre for Astrophysics Research, University of Hertfordshire, Hatfield AL10 9AB, United Kingdom\\
$^{3}$UK Network for Bridging Disciplines of Galactic Chemical Evolution (BRIDGCE), \url{https://www.bridgce.ac.uk}\\
$^{4}$Center for Gravitational Physics, Yukawa Institute for Theoretical Physics, Kyoto University, Kyoto 606-8502, Japan\\
$^5$INAF, Osservatorio Astronomico di Trieste, I-34131 Trieste, Italy\\
$^6$Astrophysics Group, Faculty of Natural Sciences, Keele University, Keele ST5 5BG, UK\\
$^7$Kavli IPMU (WPI), University of Tokyo, Kashiwa 277-8583, Japan\\
$^8$School of Physics and Astronomy, University of Edinburgh, Edinburgh, EH9 3FD, UK
$^9$INFN, Sezione di Torino, I-10125 Torino, Italy}

\recdate{July 18, 2019}

\abst{The propagation of
uncertainties in reaction cross sections and rates of neutron-, proton-, and $\alpha$-induced reactions into the final 
isotopic abundances obtained in nucleosynthesis models is an important issue in studies of nucleosynthesis and Galactic 
Chemical Evolution. 
We developed a Monte Carlo method to allow large-scale postprocessing studies of the impact of nuclear 
uncertainties on nucleosynthesis.
Temperature-dependent rate uncertainties combining realistic experimental and theoretical 
uncertainties are used. From detailed statistical analyses uncertainties in the final abundances are derived as
probability density distributions. Furthermore, based on rate and abundance correlations an 
automated procedure identifies the most important reactions in complex flow patterns from superposition of many zones or tracers.
The method so far was already applied to a number of nucleosynthesis processes. Here we focus on the production of $p$ nuclei in 
white dwarfs exploding as 
thermonuclear (type Ia) supernovae. We find generally small uncertainties in the final abundances despite of the dominance of 
theoretical nuclear uncertainties. A separate analysis of low- and high-density regions indicates that the total uncertainties are 
dominated by the high-density regions.
}

\kword{nucleosynthesis, thermonuclear supernovae, $\gamma$-process, $p$-process, nuclear reactions, Monte Carlo}

\begin{document}
\maketitle

\section{Introduction}
\label{sec:intro}

Low-energy reaction cross sections with light projectiles are required to determine astrophysical reaction rates and to constrain 
the production of 
nuclides in various astrophysical environments. Even along stability not all rates can be constrained experimentally and 
combinations of experimental data and nuclear theory have to be used. Furthermore, off stability only theoretically predicted 
reaction rates are used in nucleosynthesis calculations, both for neutron-rich and proton-rich nuclides. Our studies 
address an important question in the context of astrophysical applications: how 
uncertainties in reaction cross sections and rates of neutron-, proton-, and $\alpha$-induced reactions propagate into the final isotopic abundances obtained in nucleosynthesis models. 
This information is important for astronomers to interpret their observational data, for groups studying the enrichment of the 
Galaxy 
over time with heavy elements, and in general for disentangling uncertainties in nuclear physics from those in the astrophysical 
modelling. We developed a Monte Carlo (MC) method to allow large-scale studies of the impact of nuclear 
uncertainties on nucleosynthesis.
The MC framework \textsc{PizBuin} can perform postprocessing of astrophysical trajectories with a large reaction network, 
accounting for several thousand nuclides with several tens of thousand reactions. The trajectories specify the temporal 
evolution of density and temperature and can be taken from any astrophysical simulation. Furthermore, the analysis can be 
performed combining many such trajectories, for example, for trajectories describing different regions of an exploding star.

The method so far was already applied to a number of processes: the 
$\gamma$ process ($p$ process) \cite{tommy} and the $\nu p$ process \cite{mcnup} in core-collapse supernovae (ccSN), 
the production of $p$ nuclei in white dwarfs exploding as 
thermonuclear (type Ia) supernovae \cite{snIa} (SNIa), the weak $s$ process in massive stars \cite{nobuya}, and the main $s$ 
process in 
AGB stars \cite{gabriele}. 
Especially the studies of nucleosynthesis in thermonuclear supernovae and for the $\nu p$ process were computationally very 
demanding and 
necessitated the use of the HPC system DiRAC in the UK.

Section \ref{sec:uncert} explains how uncertainties are defined and extracted from the MC calculations. The method to 
identify key reactions is presented in Sec.\ \ref{sec:keyrates}. As an example for the application of our method, results for the 
production of $p$ nuclides in SNIa are shown in Sec.\ \ref{sec:results}.

\section{Uncertainties}
\label{sec:uncert}

Astrophysical reaction rates have to consider the thermal excitation of nuclei in the stellar plasma. They include reactions on 
target nuclei in the ground state (g.s.) and in excited states, depending on the plasma temperature. In the MC variations,
temperature-dependent rate uncertainties combining realistic experimental and theoretical 
uncertainties are used. This is necessary because experiments often only constrain ground state (g.s.) contributions to the stellar 
rates when studying reactions on nuclei above Fe. 

An uncertainty factor $U_{\mathrm{g.s.}}$ for the reaction cross 
section of a target nucleus in the g.s.\ and a factor $U_\mathrm{exc}$  for the prediction of reactions with the target 
nucleus being excited can be combined to a total uncertainty $u^*$ of the stellar rate \cite{stellarerrors,advances}, with
\begin{equation}
u^*(T)=U_{\mathrm{g.s.}}+(U_\mathrm{exc}-U_{\mathrm{g.s.}}
)(1-X_0(T)) \label{eq:uncertainty} \quad,
\end{equation}
and assuming $U_\mathrm{exc}>U_{\mathrm{g.s.}}$. The g.s.\ contribution to the stellar rate is given by
\begin{equation}
\label{eq:xfactor}
X_0= \frac{2J_0+1}{G(T)} \frac{\langle \sigma^\mathrm{g.s.}_{Aa} v \rangle}{\langle \sigma^*_{Aa} v \rangle}\quad ,
\end{equation}
with $\sigma^\mathrm{g.s.}$ being the reaction cross section for the target nucleus in the g.s.\ and $J_0$ is its g.s.\ spin. The 
quantity in pointed brackets is the reactivity, obtained by averaging the cross sections over Maxwell-Boltzmann distributed 
velocities $v$. The stellar reactivity $\langle \sigma^*_{Aa} v \rangle$ includes reactions on the g.s.\ and all thermally excited 
states. The temperature-dependent nuclear partition function is denoted by $G(T)$.

Bespoke values for $u^*(T)$ were assigned to each specific reaction and in each of the 10000 MC runs a random variation of a 
standard stellar rate $r_{\mathrm{std}}^*(T)$ was chosen by taking the rate $r^*$ used in the calculation from the interval 
$\left[r_{\mathrm{std}}^*(T)/u^*(T),r_{\mathrm{std}}^*(T)u^*(T)\right]$. Experimental uncertainties were used for 
$U_{\mathrm{g.s.}}$, where available. Theory uncertainties for $U_{\mathrm{g.s.}}$ and $U_\mathrm{exc}$ were estimated based on the 
averaged width dominating a reaction cross section.
These were motivated by comparisons between experimental and predicted g.s.\ cross sections across the nuclear chart and ranged 
from a factor of two uncertainty for neutron captures to a factor of 10 for reactions involving $\alpha$ particles \cite{tommy}.

The final abundance uncertainties then are given as probability density distributions, counting how frequently an abundance value 
appears in the simultaneous variation of all rates. The probability distributions are asymmetric and approximate a lognormal 
distribution because they result from the combined variations of many different rates.

\section{Key rates}
\label{sec:keyrates}

An automated procedure to identify the most important reactions in complex flow patterns from superposition of many zones or 
tracers was used. The method is superior to visual inspection of flows and manual variation of limited rate sets. It is based 
on examining correlations between variations in rates and abundances. A number of correlation definitions are found in literature 
and can be categorized into rank methods and product-moment methods \cite{kendall55,mathguru}. Rank correlation methods, although 
formally assumed to better account for data outliers, are losing
information in the ranking procedure and are rather unsuited for the purpose of correlating reactions and abundances.
Therefore the more suitable Pearson product-moment correlation coefficient was
used to quantify correlations \cite{pearson}. This is quite safe because data outliers to which the Pearson coefficient would be 
vulnerable do not appear in an
analytic variation of reaction rates. Moreover, it is simpler to handle, especially when calculating a combined, weighted 
correlation including many trajectories.
Positive values of the Pearson coefficients $-1\leq r\leq 1$ 
indicate a direct correlation between rate change and abundance change, whereas negative values signify an inverse correlation, 
i.e., the abundance decreases when the rate is increased. Larger absolute values $|r|$ indicate a stronger correlation and this can 
be used for extracting the most important reactions from the MC data. We define a key rate (i.e., a rate dominating the final 
uncertainty in the production of a nuclide)
by having $|r|\geq 0.65$.

\begin{figure}[tb]
\includegraphics[width=\columnwidth]{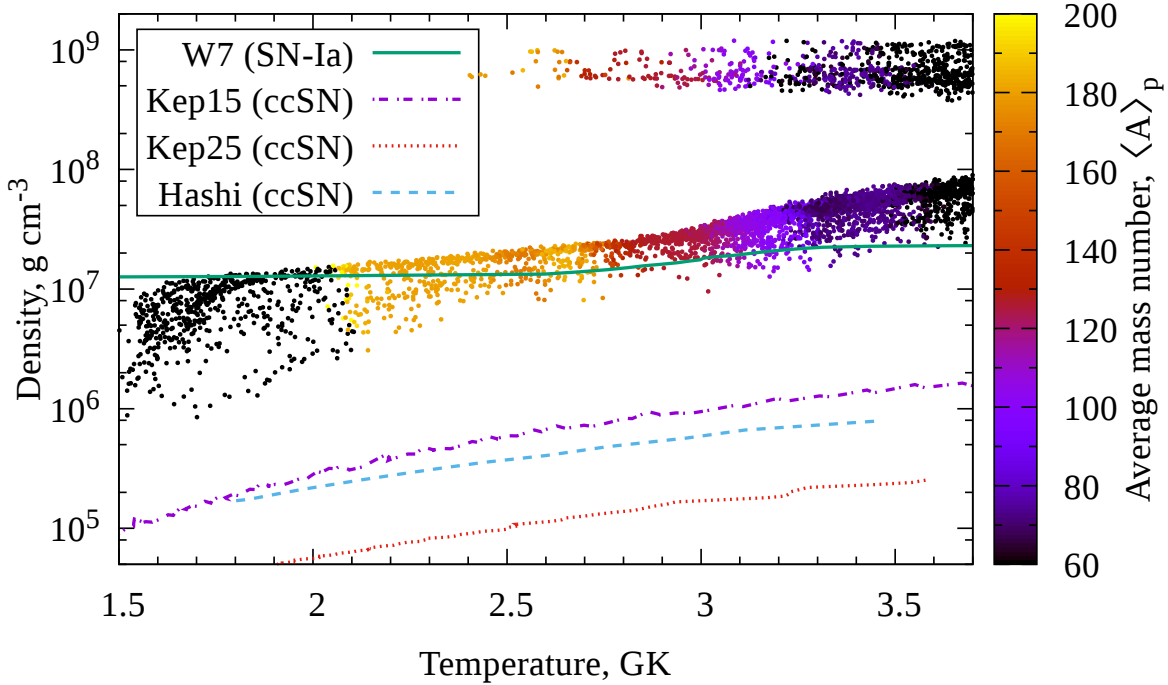}
\caption{Peak density and temperature, as well as average mass number of the produced nuclei, for $p$-nucleus producing tracers 
from a 2D thermonuclear supernova. Also shown are temperature and density conditions for 3 other explosion models; see 
text and \cite{snIa} for further details. (Figure taken from \cite{snIa}, with permission.)}
\label{fig:f1}
\end{figure}

\begin{figure}[tb]
\includegraphics[width=\columnwidth]{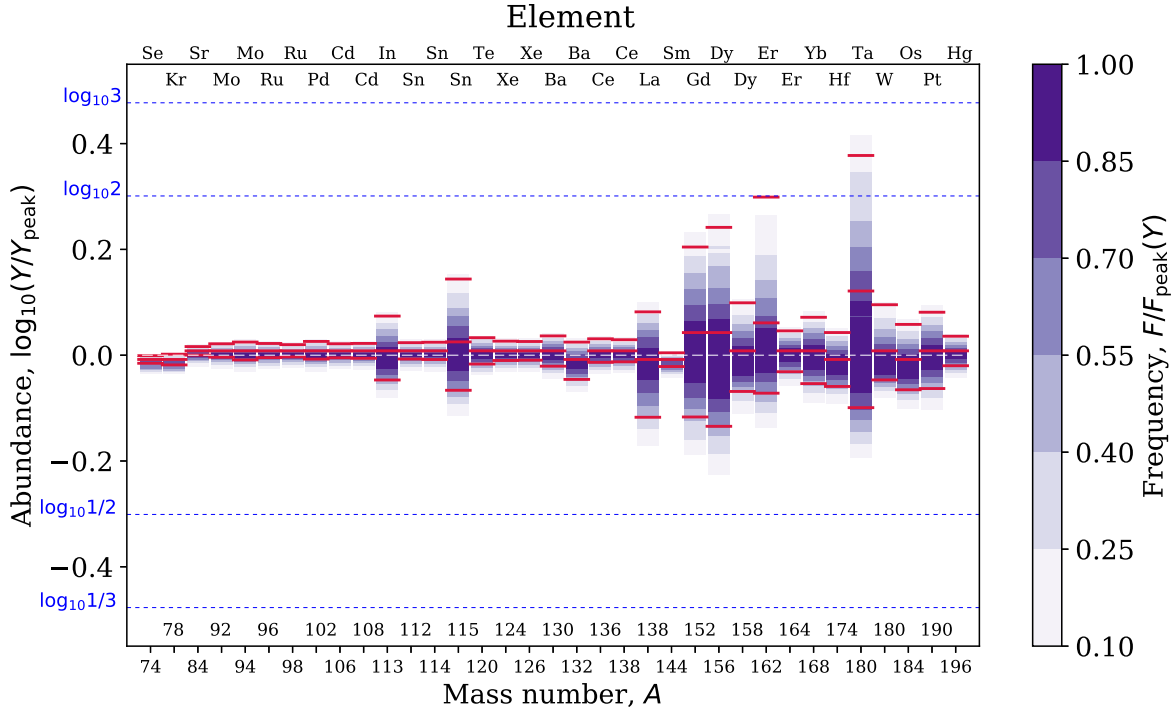}
\caption{Uncertainties in the final abundances caused by rate uncertainties. The two outer red lines in each distribution 
encompass a 90\% confidence interval. The third red line in between the outer red lines marks the 50\% level of the summed 
probability. It does 
not necessarily coincide with the most probable abundance value at the darkest color shade. (Figure taken from \cite{snIa}, with 
permission.)}
\label{fig:uncertall}
\end{figure}

\begin{figure}[tb]
\includegraphics[width=\columnwidth]{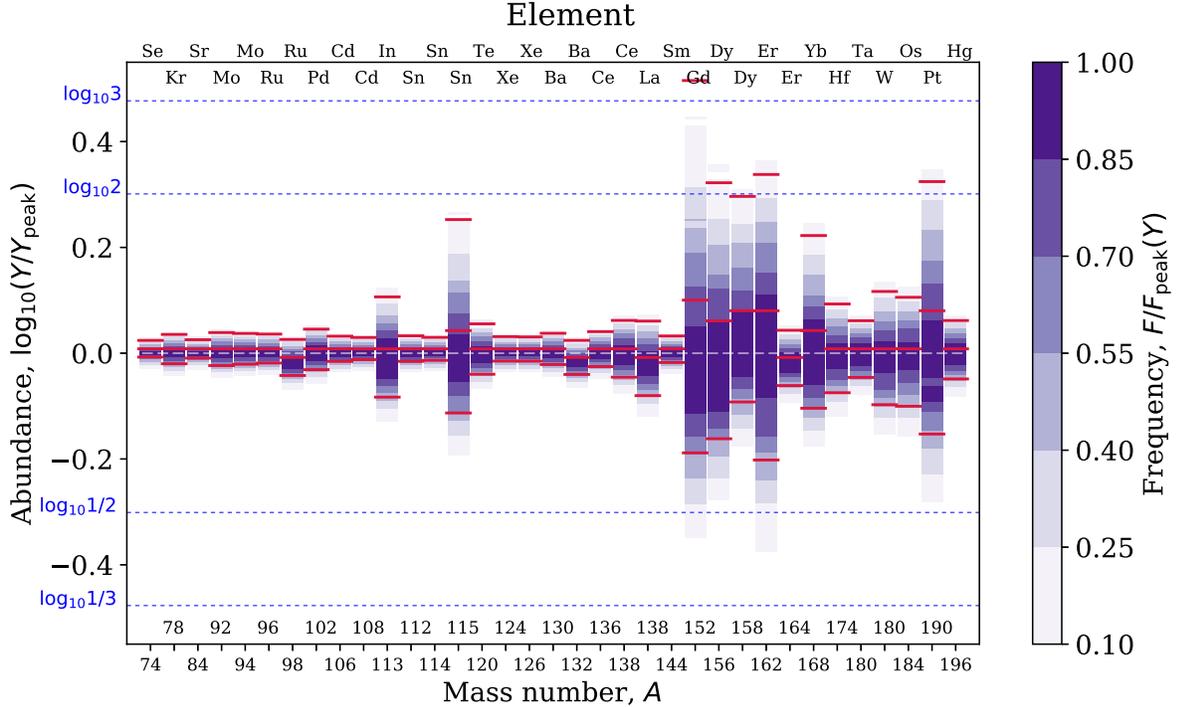}
\caption{Same as Fig.\ \ref{fig:uncertall} but only considering high-density regions. (Figure taken from \cite{snIa}, with 
permission.)}
\label{fig:uncerthi}
\end{figure}

\section{Production of $p$ nuclides in thermonuclear supernovae}
\label{sec:results}

Thermonuclear supernovae (single-degenerate type Ia supernovae) originating from the explosion of a white dwarf accreting mass from 
a companion star have been suggested as
a site for the production of $p$ nuclides \cite{trav11}. In the explosion, the photodisintegration of nuclei can 
create proton-rich nuclides, similar to a $\gamma$-process in ccSN \cite{tommy,snIa}. The prerequisite 
for production of $p$ nuclides is the existence of the single-degenerate path to a SNIa and a pre-explosive production of seed 
nuclei during H/He-burning on the surface of the white dwarf.

In this study, 51200 tracer particles were extracted from a 2D explosion simulation as described in \cite{trav11} (the model DDT-a 
was 
used here). Among these, 4624 tracers experienced conditions supporting the production of $p$ nuclides and their temperature and 
density profiles (trajectories) were used in the MC
post-processing. The reaction network included 1342 nuclides (around stability and towards the proton-rich 
side).
To process all relevant trajectories the full reaction network had to 
be run several $10^7$ times.

Figure \ref{fig:f1} shows the tracer conditions and also shows the average mass number of the produced
nuclides. A comparison to conditions in the standard W7 model of an SNIa and three models of ccSN is 
also given. While the temperature range is similar in all models -- because it is required for the production of $p$ nuclides 
and is set by the binding energies of the nuclides -- 
thermonuclear supernovae produce $p$ nuclei at considerably higher density than ccSN. It can also be seen that the tracers from the 
2D model fall in two density groups. The high-density group originates from the inner part of the white dwarf whereas the tracers 
with lower density are from the surface region.

The probability density distributions for the final abundances from the combination of all tracers are shown in Fig.\ 
\ref{fig:uncertall}.
The uncertainties are well below a factor of two, with the exception of
$^{180}$Ta. The states $^{180g}$Ta and $^{180m}$Ta were not followed separately in the network and thus the uncertainty in
$^{180m}$Ta production may be even larger, depending on the unknown, actual equilibration between g.s.\ and isomeric state. 
It can be noted that the uncertainties are generally larger from $^{152}$Gd on to higher mass numbers, with
$^{162}$Er approaching an uncertainty of a factor of two at the upper limit. Incidentally, this is also the nuclear mass region
where the calculations by \cite{trav11} exhibited problems to reproduce the solar abundance pattern for several nuclides.
It should further be noted that $^{113}$In, $^{115}$Sn, $^{138}$La, $^{164}$Er, $^{152}$Gd, and $^{180m}$Ta receive 
strong contributions from other processes than the $\gamma$ process \cite{trav11,p-review}. Therefore these 
nuclides may not be viewed as pure $p$ nuclides.

The generally low uncertainty level can be understood by the fact that the SN Ia trajectories cover a wider range of
density-temperature combinations than, e.g., those of O/Ne layers in ccSN (see Fig.\ \ref{fig:f1}). This leads to a more
complicated flow pattern without a well defined ``path'' when combining all trajectories. In such a flow, variations of a few rates
do not greatly affect the general flow as they can easily be bypassed by other reactions, thus avoiding an
abundance variation.

It was found that the uncertainties in the final production are dominated by the uncertainties 
in the high density group, which are shown in Fig.\ \ref{fig:uncerthi}. Even in this tracer group, the uncertainties stay well 
below a factor of two, except for isotopes of Gd, Dy, Er, and Pt.

There were no key rates found dominating the production of the $p$
nuclides, indicating that the uncertainties in several rates conspire to produce the shown total uncertainties. The only key 
rate found in this study was $^{145}$Eu(p,$\gamma$)$^{146}$Gd, affecting the production of $^{146}$Sm. The lack of further key 
rates indicates complex flow patterns with several different production paths for most of the nuclides.

\section{Conclusion}

A comprehensive, large-scale MC study of the production of $p$ nuclei was performed using more than 4000 trajectories from a 2D 
model of a single-degenerate, thermonuclear supernova. Astrophysical reaction rates on several thousand target nuclides were 
simultaneously varied within individual temperature-dependent uncertainty ranges constructed from a combination of experimental and 
theoretical error bars. This allowed to investigate the combined effect of rate uncertainties leading to total uncertainties in the 
final abundances.

The total uncertainties 
found were well below a factor of two, despite of the contribution of a large number of unmeasured rates with large theory 
uncertainties. Tracers with high density conditions, originating from the inner part of the white dwarf, contribute strongly to the 
final, total uncertainties.

The tracer distribution used here may be specific to a certain 2D explosion model. Other simulations, especially 3D models, may 
lead to a different relative weight in the number of regions with low or high density. Our conclusions remain generally valid, 
though, due to the separate analysis of low- and high-density tracers. Using the present results, uncertainties in the production 
of $p$ nuclides in other models can also be assessed by combining the uncertainties derived for the two density groups.

In conclusion, we found that the uncertainties stemming from uncertainties in the astrophysical reaction rates are small compared to 
the uncertainties arising from the choice of site, explosion model, and numerical treatment of the explosion hydrodynamics.

\section*{Acknowledgments}
This work has been partially supported by the European Research Council 
(EU-FP7-ERC-2012-St Grant 306901, EU-FP7 Adv Grant GA321263-FISH), the EU COST Action CA16117 (ChETEC), the UK STFC (ST/M000958/1), 
and MEXT Japan (``Priority Issue on Post-K computer: Elucidation of the Fundamental Laws and Evolution of the Universe'' and ``the 
World Premier International Research Centre Initiative: WPI Initiative''). G.C. acknowledges financial support from the EU 
Horizon2020 programme under the Marie Sk\l odowska-Curie grant 664931. Parts of the computations were carried out on 
COSMOS (STFC DiRAC Facility) at DAMTP in University of Cambridge. This equipment was funded by BIS National E-infrastructure capital 
grant ST/J005673/1, STFC capital grant ST/H008586/1, and STFC DiRAC Operations grant ST/K00333X/1. DiRAC is part of the UK National 
E-Infrastructure. Further computations were carried out at CfCA, National Astronomical Observatory of Japan, and at YITP, Kyoto 
University. The University of Edinburgh is a charitable body, registered in Scotland, with Registration No.~SC005336.

\end{document}